\documentclass[12pt,tightenlines,
nofootinbib,pra, twocolumn, showpacs,preprintnumbers]{revtex4}

\usepackage{epsfig}

\newcommand{\beq}{\begin{equation}}
\newcommand{\eeq}{\end{equation}}
\newcommand{\bea}{\begin{eqnarray}}
\newcommand{\eea}{\end{eqnarray}}

\def\d{\partial}
\def\<{\langle}
\def\>{\rangle}

\begin{document}

\title{  Vortex Rings in two Component Bose-Einstein Condensates}
\author{Max~A.~Metlitski and Ariel~R.~Zhitnitsky}
\affiliation{Department of Physics and Astronomy, \\
University of British Columbia, \\
Vancouver, BC V6T 1Z1, Canada  }

\begin{abstract}
We study the structure of  the vortex core  in two-component Bose-Einstein
condensates.
We demonstrate that   the order parameter may not vanish and  the symmetry may
not be restored in the core of the vortex. In this case such vortices can form
 vortex rings known as   vortons  in particle physics literature. In contrast
with well-studied superfluid $^4He$, where similar vortex rings
can be stable due to Magnus force only if they move, the vortex
rings in two-component BECs can be stable even if they are at
rest. This beautiful effect was first discussed by Witten in the
 cosmic string context, where it was shown that the
 stabilization occurs due to condensation of the second component of the
field
 in the
vortex core. This second condensate trapped in the core may carry
a current along the vortex ring counteracting the effect of string
tension that causes the loop to shrink. We speculate that such
vortons may have been already observed in the laboratory. We also
speculate that the experimental study of topological structures in
BECs can provide a unique opportunity to study cosmology and
astrophysics by doing laboratory experiments.

\end{abstract}

\maketitle

\section{Introduction}
 With the recent controlled creation of vortices
\cite{BoseVortex_1,BoseVortex_1a} in trapped atomic Bose-Einstein
condensates (BECs), the study of structure and dynamics of
topological defects in weakly interacting superfluids has become
an active subject of experimental and theoretical research.

Similar topological structures in form of vortex lines (strings)
have been observed in superfluid $^3He$ and $^4He$,
 conventional superconductors and  high $T_c$ superconductors. In most
studied and well understood cases such as the $^4He$ system and
the conventional superconductors, the order parameter can be
represented by a single complex scalar field, and the structure of
the vortex is quite simple. It can be understood in terms of
spontaneous violation of the
   $U(1)$ symmetry, which represents the conservation of number of particles
(in $^4He$) or electric charge (in conventional superconductors).
As is well-known, spontaneous violation of the $U(1)$ symmetry
 leads to a non-zero magnitude of the  order parameter, which represents
 the
superfluid density (in $^4He$)  or the Cooper pair density (in
conventional superconductors). The structure of the vortex core in
these cases is trivial:  the order parameter must vanish in the
center of the core, restoring the $U(1)$ symmetry there, see e.g.
\cite{defectsbook}. If one makes a ring out of such a vortex, it
would quickly shrink and decay due to the large string tension.
The well-known quantized vortex rings, which have been produced
and detected \cite{Rayfield} in superfluid $^4He$, were stable due
to their motion that gives rise to a Magnus force equilibrating
the string tension. We shall not discuss vortex ring stabilization
by Magnus forces in the current work.

The subject of this work is the analysis of a less trivial and
much more interesting situation when the vortex rings, dubbed
vortons,
 can be stabilized even when they do not move. As we discuss below,
such a stability cannot occur in a system described by a single
scalar field; rather such a stability  may occur in more complex
systems, such as the two-component BECs, which are considered in
this paper.

We should note at this point that the theoretical feasibility for
such  a phenomenon to occur, was first discussed by E. Witten who
suggested a toy model leading to existence of superconducting
cosmic strings \cite{Witten}.  The idea of construction is simple:
consider a two component system (such as a two component BEC),
which is  described by two complex scalar fields $(\psi,\phi)$
with an approximate
  $U(2)$ symmetry. If the $U(2)$ symmetry between fields $\psi$ and
$\phi$ is explicitly  broken down to $U(1) \times U(1)$, the
$\psi$ condensation might be energetically more favorable than
$\phi$ condensation, and the ground state will be given by
$\<\psi\> \neq 0$ and $\<\phi\> = 0$. As is well known, such a
system allows for existence of $\psi$ vortices characterized by
the phase of $\psi$ field varying by an integer multiple of $2
\pi$ as one traverses a contour around the vortex core. What
Witten actually has demonstrated is that if the approximate $U(2)$
symmetry is broken only weakly, the $\phi$ field may condense
inside the core of the $\psi$ string, "spontaneously breaking" the
corresponding $U(1)$ symmetry in the core. In the case when the
$\phi$ field is electrically charged, such a string can support a
persistent electric current along the string core, hence the term
"superconducting cosmic strings".
 As will be discussed below, the question of existence and magnitude
of $\phi$ field condensation in the vortex core depends solely on
how strongly the $U(2)$ symmetry is broken.

One should remark here that the above idea (that the core of the
string could be in a different phase without restoration of
symmetry) was motivated by the grand unified theories with a
typical scale of $10^{15} GeV$. However, this phenomenon was
experimentally observed in a very different environment with a
typical scale of $10^{-4} eV$. Here we are referring to the
experimental observation of ferromagnetic $^3He-A$ cores in
superfluid $^3He-B$ vortices, for review see
\cite{VolovikBook,Volovik}. Also, this phenomenon has been
predicted to occur in the SO(5) model of high-temperature
superconductivity, where the cores of the conventional Abrikosov
strings are in the
  antiferromagnetic state \cite{Zhang,SO5Vortons}.
There has been recent experimental evidence that suggests this
theoretical picture may be correct
\cite{vaknin,lake,dai,mitrovic,miller}. It has been argued
\cite{Kaplan,KStrings, KVortons} that this phenomenon may also
occur in the so-called color superconducting phase of QCD that is
believed to be realized when the baryon density is a few times
larger than nuclear density \cite{CFL}.

The consequences of core-condensation are far-reaching and cannot
be overseen at this point. In particular, as will be demonstrated
below, one expected consequence of core-condensation is expansion
of the vortex core as the magnitude of explicit $U(2)$ symmetry
breaking decreases. Thus, in the limit of very weak $U(2)$
symmetry breaking the vortex cores in 2-component BECs may become
large enough to allow for direct experimental study of core
properties. One other known consequence of core condensation is
that the interactions between vortices can be drastically altered
by the presence of non-trivial cores \cite{MacKenzie}. If this
indeed happens, the standard picture of Abrikosov lattices in high
$T_c$ superconductors or vortex lattices in BECs may not be
necessarily always correct.

Another consequence of the phenomenon of core-condensation was
discussed previously  in the context of cosmic strings. Within
Witten's original toy model, it has been shown \cite{Turner,
Hindmarsh,Davis, DavisShell1, DavisShell2, DavisShell3,
DavisShell4, Shellard_num} that
condensation inside the core provides a way to stabilize a string
loop against shrinking, see \cite{Shellard2002} for review.
 Indeed, a string loop is a
topologically trivial object that normally shrinks due to string
tension. However, if one allows for the $\phi$ field condensed in
the core of a $\psi$ vortex to carry a winding number along  the
string as well as a Noether's charge associated with its $U(1)$
symmetry,
 then the conservation
of both (the  charge and the winding number)  prevents the loop from shrinking.
 In fact, when both Noether's charge  and winding
number are present, the $\phi$ field carries a non-zero angular
momentum perpendicular to the plane of the loop, which stabilizes
the configuration. The resulting stable vortex loops, known as
vortons, are called semi-topological defects since they are
stabilized in part by topology and in part by energetics. One
should comment here that this stability is purely mechanical in
origin and to a leading approximation is independent of whether
the relevant symmetry is local or global.

Up to this point, most discussion of vortons has been confined to
relativistic physics (such as cosmic strings and high-density
QCD), or to physics where the relevant low-energy effective
Lagrangian has a relativistic form (such as the SO(5) model of
high $T_c$ superconductivity). In the present work we wish to
investigate whether counterparts of semi-topological defects
described above exist in non-relativistic field theory, such as
the one describing a mixture of two ultra cold Bose condensed
gases. In this paper we discuss the structure of BECs with two
internal levels. This is equivalent to a spin-1/2 fluid: the order
parameter has $U(2)$ rotational properties. We demonstrate that
strings with condensation in the core do exist if the $U(2)$
symmetry is slightly broken, and we calculate explicitly under
which conditions this occurs. We then go on to construct vortons
in non-relativistic systems and demonstrate their stability under
certain conditions.

The application to the field of Bose gases seems particularly
interesting in the wake of recent theoretical and experimental
achievements in this area. On the experimental side, the long
sought goal of obtaining a weakly interacting Bose gas is now
achieved in ultra-cold alkali metal gases such as $Rb$, $Na$ and
$Li$ \cite{Rb,Na,Li}. Moreover, experimentally, it is now possible
\cite{Bose_2_Comp1, Bose_2_Comp2} to have a system with two
  hyperfine states of $^{87}Rb$ condensed in the same trap;
 this is precisely the kind of system we consider in the present work.
 Also, as we already mentioned, in the past few years
experiments have confirmed that rotational vortices do indeed form
in the ultra-cold Bose gases
\cite{BoseVortex_1,BoseVortex_1a,BoseVortex_2}. Actually, the
original experiment that produced a quantum vortex in $^{87}Rb$
\cite{BoseVortex_1} for technical reasons made use of a second
condensate in the core of the vortex. Thus, we speculate that the
phenomenon of core condensation described in this paper might have
been already observed in BECs. However, since in the original
experiment the second field component was artificially induced in
the core, one cannot immediately identify the phenomenon of
natural core condensation that we are discussing with the
experimentally observed result. Moreover, we note that vortex
rings, which are the subject of the present work, may have also
been observed \cite{ring}. However, an additional analysis is
required before one can convincingly argue that the observed rings
are precisely vortons, which are stable due to the condensation in
the core rather than due to some other factors such as nonzero
velocity of the ring or due to the influence of the trap
potential.

On the theoretical side, the
weakly interacting BECs provide a fascinating tool for
testing theoretical ideas in various fields where
experimental control is not possible (such as cosmology and
astrophysics, see some comments in the Conclusion).
 Recently there has been a number of theoretical
investigations into topological defects such as domain walls,
vortices and skyrmions in Bose-Einstein condensates
\cite{Mueller}-\cite{Battye}. Most of these papers, however, are
focused on the case when only a single field is present so that
the effects discussed in these paper do not occur at all. In other
papers\cite{Mueller,Son} two particle species are present in the
ground state. In this case, the effect that is the subject of this
paper (when the second condensate is suppressed in the bulk of the
media but reappears in the core of the vortex) is not as
pronounced. In \cite{Kinks} a configuration known as a
"dark-bright vector soliton" is discussed, consisting of a domain
wall formed by one condensate with the second condensate confined
to the wall's center. This defect is just a 1-dimensional analogue
of the vortex with a core condensate considered in this paper. In
other related works\cite{Anglin,Battye,Stoof,Ruostekoski} a
nontrivial vortex ring in a two-component BEC is interpreted as
the skyrmion, a topologically stable solution originally
introduced in particle physics literature\cite{Skyrme}. While the
analysis\cite{Anglin,Battye,Stoof,Ruostekoski} is mainly
numerical, and therefore, it is quite difficult to make a precise
correspondence with our analytical approach, we have a strong
feeling that configurations described in
\cite{Anglin,Battye,Stoof,Ruostekoski} are more similar to
non-topological vortons considered in the present work rather than
to topological skyrmions as claimed, see arguments in
\cite{skyrmion}. The advantage of the analytical approach
advocated in the present work is the clear understanding  of the
structure of the solution as well as the nature of the stability
of the configuration. Numerical calculations are more useful when
the influence of the trapping potential (and other complications,
which are always present in real experiments) cannot be neglected
and should be taken into account.

This paper is organized as follows. In Section II we review the
properties of the conventional $U(1)$ strings and construct explicit numeric and
variational solutions for the $\psi$ string. In Section III, we
determine the precise conditions under which string core
condensation occurs and construct numerical solutions for $\phi$
and $\psi$ fields in this regime. In Section IV, we describe
vortons in non-relativistic physics and demonstrate their
classical stability.

\section{The Lagrangian and the conventional $U(1)$ strings}
We wish to consider a non-relativistic system with two types of
particles such that two different species are very similar to each
other and the number of particles of each type is conserved, i.e.
the exact symmetry of the system is $U(1) \times \tilde{U}(1)$. We
also assume that the $s$ wave scattering lengths between like and
unlike species are numerically close to each other, such that the
relevant symmetry is even larger, as will be explained below.
Finally, we assume here that the trapping potentials are
sufficiently wide so that in what follows we neglect all the
boundary effects. The last simplification is mainly motivated by
our interest in effects, which  may occur in very large  systems
(such as the systems, which appear in  cosmological  and
astrophysical context) rather than in finite volume systems where
the trapping potential plays an essential role.
 Thus, throughout
this paper, we consider the following Lagrangian
  describing a
system of two weakly interacting ultra-cold Bose-gases of same
mass:
\begin{widetext}
\beq \label{DimLag} {\cal L} = i \hbar {\tilde{\Phi}}^{\dagger}
\tilde{\d}_0 \tilde{\Phi} - \frac{\hbar^2}{2 m} |\tilde{\nabla}
\tilde{\Phi}|^2 + \mu_1 |\tilde{\psi}|^2 + \mu_2 |\tilde{\phi}|^2
- \frac{1}{2} g_{11}|\tilde{\psi}|^4 - \frac{1}{2}
g_{22}|\tilde{\phi}|^4 - g_{12}|\tilde{\psi}|^2 |\tilde{\phi}|^2
\eeq
\end{widetext}
where $\tilde{\Phi} = (\tilde{\psi}, \tilde{\phi})$ is a doublet
of complex scalar fields\footnote{The original dimensionfull fields and
space-time variables appearing in (\ref{DimLag}) carry a sign
"tilde". Throughout this paper we will be mostly using a
dimensionless notation for the fields and space-time variables,
which will be introduced below; these will carry no sign
"tilde".}. The chemical potentials of particles ($\tilde{\psi}$,
$\tilde{\phi}$) are $(\mu_1, \mu_2)$ and coupling constants
$g_{ij}$ are proportional to scattering lengths $a_{ij}$ via: $
g_{ij} = \frac{4 \pi \hbar^2}{m} a_{ij}$.
  By varying the action $S=\int{\cal
L}d^3\tilde{x}d\tilde{t} $ with respect to $\tilde{\psi},
\tilde{\phi}$, one can derive the familiar Gross-Pitaevskii
equations. 


If $(\tilde{\psi},\tilde{\phi})$ denote two hyperfine states of
$^{87} Rb$, the scattering lengths $a_{11}, a_{12}, a_{22}$ differ
by only about $3\%$.
 We neglect this difference
in what follows, and assume $a_{ij} = a$, $g_{ij} = g$, such that
the Lagrangian (\ref{DimLag}) can be written in a $U(2)$ notation
as follows,
\begin{widetext}
\beq \label{DimLagMod} {\cal L} = i \hbar \tilde{\Phi}^{\dagger}
\tilde{\d_0} \tilde{\Phi} - \frac{\hbar^2}{2 m} |\tilde{\nabla}
\tilde{\Phi}|^2 - \frac{g}{2}(|\tilde{\Phi}|^2 -
\frac{\mu_1}{g})^2 + (\mu_2 - \mu_1) |\tilde{\phi}|^2 \eeq
\end{widetext}
In the present work we assume $\mu_1 > \mu_2$, and hence the
$\tilde{\psi}$ field will condense, while the $\tilde{\phi}$ field
will remain uncondensed in the ground state.

It proves to be very convenient to introduce a dimensionless
notation for fields, as well as for spacial (and time)
coordinates. In the dimensionless notation all distances are
measured in units of the correlation (healing) length $\xi=
({\hbar^2}/{2 m \mu_1})^{1/2}$, all frequencies are measured in
units $\mu_1/\hbar$, while the absolute values of the fields $
|\tilde{\phi}|^2, |\tilde{\psi}|^2 $ are measured in units of
particle density
 $n= {\mu_1}/{g}$, see Appendix for precise correspondence.
Throughout this paper, we use the sign  "tilde" to distinguish
dimensionless (without tilde) and original, dimensionfull (with
tilde) variables. Although we will be primarily using the
dimensionless notation, we will sometimes restore all physical
units in key formulas; we hope this will not confuse the reader.
 With all these
remarks in mind, our starting point is the following dimensionless
Lagrangian,
\beq
\label{Lagrangian}
 {\cal L} = i \Phi^{\dagger}\d_0 \Phi - {\cal H}, \qquad \Phi = (\psi,\phi) .
\eeq Here the Hamiltonian density ${\cal H}(\psi, \phi)$ has the
form, \beq\label{H} {\cal H} =  |\nabla \Phi|^2 +
\frac{1}{2}(|\Phi|^2 - 1)^2 + \delta m^2 |\phi|^2, \eeq where all
derivatives are with respect to the dimensionless coordinates
introduced in the Appendix. This Lagrangian is exactly equivalent
to eq.(\ref{DimLagMod}). It now is apparent that the only relevant
parameter entering the Hamiltonian is the difference in chemical
potentials of $\psi$ and $\phi$ fields, which is parameterized by
a dimensionless factor $\delta m^2\equiv ({\mu_1-\mu_2})/{\mu_1}$.
In expression (\ref{Lagrangian}), $\Phi$ is a doublet of complex
scalar fields, and the Lagrangian
  possesses an exact  $U(1) \times \tilde{U}(1)$ symmetry
with respect to independent phase rotations of the fields,
$$U(1): ~\psi\rightarrow e^{i\alpha_1}\psi, ~~\tilde{U}(1): ~\phi\rightarrow e^{i\alpha_2}\phi .$$
The corresponding conservation laws are those of the number of
particles of each specie, $N_1=\int d^3x \psi^{\dagger}\psi$ and
$N_2=\int d^3x \phi^{\dagger}\phi$. Moreover, the Lagrangian
(\ref{Lagrangian}) possesses an approximate $U(2)$ symmetry, which
is explicitly broken by the $\delta m^2 |\phi|^2$ term. In the
limit $\delta m \rightarrow 0$ the Lagrangian acquires an exact
$U(2)$ symmetry,
$$\Phi\rightarrow e^{i\alpha}e^{i\frac{\sigma^a}{2}\theta^a}\Phi ,~~~~
 Tr(\sigma^a\sigma^b)=2\delta^{ab}$$
with $\sigma^a$ being the Pauli $\sigma-$ matrices.
 All of these symmetries are global,
although local $U(1)$ symmetries could also be considered without
changing the results of this work qualitatively.
Without loss of generality, throughout this paper,
 we consider the case $\delta m^2 > 0$. Then
by minimizing the potential energy, we find that the vacuum
expectation value for $\psi$ field is non-zero, which we choose to
be  $\<\psi\>$ = 1, while   $\<\phi\>$ = 0. Thus, the $U(1)$
symmetry is spontaneously broken, while the $\tilde{U}(1)$
symmetry remains an exact symmetry.
 Hence the vacuum manifold of the system is a
circle $S^1$ with points characterized by the phase of the field
$\psi$. As we know, spontaneous breaking of a $U(1)$ symmetry
leads to formation of topological defects known as strings.
Therefore, one should expect the existence of these objects for
arbitrary non-zero values of parameter $\delta m^2$.

Let us review here the elementary properties of $U(1)$ strings,
which will be used later in the text. Until further notice, we
will consider all the fields to be uniform in the 3rd (string)
direction and, thus, constrain our attention to a plane crossing
the string at a right angle. A string is a time-independent
configuration,
 which exhibits a non-trivial mapping of the plane boundary $S^1$ into the vacuum manifold.
In particular, for the $U(1)$ string, as one traverses the circle
out at infinity in real space, the phase of the field $\psi$ makes
an integer number of windings $l$ in the vacuum manifold $S^1$.
Explicitly, \beq \label{Psi_Winding}
 l = \frac{1}{2 \pi}{\int}_{\Gamma} d \arg
(\psi (x)) \eeq
where $\Gamma$ is some large circle in the plane
crossing the string. Such configurations are absolutely stable
since it is impossible to continuously deform configurations with
different values of topological charge $l$ into each other without
leaving the vacuum manifold.
The first important feature of this configuration is that
  if $l \neq 0$, $\psi$ must vanish at
some point in order to be properly defined, see e.g.\cite{Coleman}.
 This implies that
$\psi$ leaves the vacuum manifold somewhere and the configuration
certainly possesses a non-zero energy. Our next remark is that if
the $U(2)$ symmetry between $\psi$ and $\phi$ fields was exact
(i.e. $\delta m^2$ = 0) then the vacuum manifold would have been
given by $|\Phi| = 1$, rather than $|\psi|=1$.
As is well known,
there are
 no topologically stable strings possible in this case as there is always a direction
 in the configurational space through which any
 windings of the $\psi$ field can be continuously removed.
An important consequence of the above formal fact is that in two
component BECs the only reason for the existence of vortices is an
explicit violation of the $U(2)$ symmetry. Without this violation,
the strings would immediately decay. Thus, we expect the typical
characteristics of strings such as the core size to be ultimately
related to the magnitude of violation of the $U(2)$ symmetry, see
the detailed discussion in the next section.

Let us now construct explicit variational and numerical string
configurations, which will be used later in the text.
Formally, we wish to find a solution to the following equations of
motion,
 \beq \label{Psi_Equation}
{\nabla}^2 \psi (x) = (|\psi (x)|^2 -1) \psi (x) ,~~ \phi=0. \eeq
such that the $\psi$ field satisfies the string boundary
conditions, i.e. it carries a winding number $l$. Here we have
assumed that in the lowest energy state possessing the string
boundary conditions, the $\phi$ field nowhere leaves its vacuum
expectation value (we will see in the next section that this need
not be the case). The standard cylindrically symmetric ansatz
corresponding to the string configuration is, \bea
\label{String_Ansatz} \psi(x) = f(r) e^{i l \varphi},~~~~~~~~~
 \nonumber\\
f(r \rightarrow 0) \rightarrow 0,~~ f(r \rightarrow \infty)
\rightarrow 1 \eea where $(r,\varphi)$ are the standard polar
coordinates, and $f(r)$ is real. It is instructive to estimate the
string tension (energy per unit vortex length) $\mu$ before
presenting any  numerical or variational calculations. By
definition, from eq. (\ref{H}), the string tension is,
\bea \label{mu} \mu=\int d^2x{\cal H} =  \\
\label{mupsi}
 \int d^2x\Big(|\nabla
\psi|^2 + \frac{1}{2} (|\psi|^2 -1)^2\Big).
 \eea
For large $r$, the leading contribution to $\cal H$ comes from the
gradient part associated with the phase variation of the $\psi$
field, ${\cal H} \approx l^2/r^2$, and is not sensitive to the
specifics of interactions.  Now, performing the integration
  over a disc of radius $\Lambda$, we estimate
 $\mu \approx 2 \pi l^2 \log(\Lambda)$, where $\Lambda$
is a typical distance between vortices. In physical units (see
Appendix) the string tension is given by \beq \label{mu1}
\tilde{\mu}\simeq  \frac{\hbar^2}{2m} n \cdot 2 \pi l^2
\log(\frac{\tilde{\Lambda}}{\xi}), \eeq where $n=
|\tilde{\psi}(r=\infty)|^2$ is the density of the BEC measured far
away from the vortex core.
 From this simple estimate it is clear that vortices with
winding number $l > 1$ are not stable with respect to decay into
$l$ vortices of unit winding number. Thus, throughout this paper
we focus on simple vortices with $l = 1$.

Now we want to construct an explicit variational solution to a
vortex with $l=1$ to be used later in the text. Following
\cite{Turner} we take a variational ansatz that satisfies our
boundary conditions: \beq \label{Var_Ansatz} f(r) = 1-e^{-\beta r}
 \eeq
where $\beta$ is a free variational parameter. Substituting this
into the energy density (\ref{H}), integrating over 2-dimensional
space and minimizing with respect to $\beta$ we find: \beq
\label{Var_Soln} \tilde{\beta} = \frac {\sqrt{89}}{12\xi}, ~~
\tilde{\mu} = \frac{\hbar^2}{2m} n \cdot 2\pi \Big(
\log(\tilde{\beta} \tilde{\Lambda})+\frac{3}{4} \Big) \eeq where
we restored the physical dimensional units. Thus, we notice that
in the absence of the second condensate ($\phi = 0$), the radius
of the string core is of order $\xi$ - the typical correlation
length in the theory.

In order to test our variational solution (\ref{Var_Ansatz}) we
have also solved the exact eq.(\ref{Psi_Equation}) with
appropriate boundary conditions (\ref{String_Ansatz}), \bea
\label{f_Equation} f'' + \frac{1}{r} f' - \frac{l^2}{r^2} f - (f^2
-1) f = 0
\nonumber\\
f(r \rightarrow 0) \rightarrow 0,~~ f(r \rightarrow \infty)
\rightarrow 1 \eea numerically. We find an excellent agreement
between our variational and numerical solutions.  This gives us
confidence that one can use either of these solutions for a more
complicated problem when the configuration is unstable with
respect to formation of a $\phi$ field condensate in the core of
the $\psi$ vortex, which is the subject of the next section.

%
%
\section{Condensation in the String Core}
As we know from
   the previous section, if
  the parameter $\delta m^2$   is zero,
the Lagrangian is invariant under the symmetry group $SU(2) \times
U(1)  \rightarrow U(1)$ (broken down to $U(1)$). From topological
arguments we know that such a Lagrangian does not allow for
existence of vortices since the vacuum manifold is a 3-sphere and,
therefore, does not have non-contractible loops. In the opposite
limit, when
  $\delta m^2$ is relatively large,
the residual symmetry group is $U(1) \times \tilde{U}(1)
\rightarrow \tilde{U}(1)$ and the vacuum manifold is a circle
$S^1$, leading to formation of stable vortex solutions. From these
two limiting cases, it is clear that there should be some
intermediate region that somehow interpolates (as a function of
$\delta m^2$) between the two cases. Here is a possible scenario,
which as we shall see in a moment turns out to be correct. For
large values of $\delta m^2$ nothing interesting happens:
conventional vortices of $\psi$ field do exist, the symmetry is
restored in the string core, the core size is fixed by the
standard correlation length $\xi$ and the $\phi$ field vanishes
everywhere in space. However, when $\delta m^2$ starts to
decrease, at some finite magnitude of $\delta m^2$, an instability
arises through the condensation of $\phi$ field inside the vortex
core. As the magnitude of $\delta m^2$ further decreases, the
strength of the $\phi$ condensate in the core grows and the size
of the core, where both $\phi$ and $\psi$ are far away from their
vacuum expectation values, becomes larger and larger. Finally, at
$\delta m^2=0$ the core of the string fills the entire space, in
which case the meaning of the string is completely lost, and we
are left with the situation when the $U(2)$ symmetry of the
Lagrangian is exact: no stable strings are possible as we know
from topological arguments.

Actually, it is of little surprise that there is a tendency for
the $\phi$ field to condense in the vortex core. Indeed, if $0 <
\delta m^2 < 1$, the potential energy of the system has a global
minimum at $\psi \neq 0, \phi = 0$, and a local minimum at $\psi =
0, \phi \neq 0$. Thus, the vacuum structure of the theory is given
by condensation of $\psi$ rather than $\phi$. Yet recall that in
the core of the vortex $\psi$ must vanish, and thus the system is
driven in the core towards the local minimum where the $\phi$
field condenses. However, the energetic benefit of condensing
$\phi$ in the core competes with the tensional energy needed for
$\phi$ to relax back to its $0$ vacuum expectation value at
infinity. As was hinted in the previous section, the question of
whether $\phi$ indeed condenses in the core is determined by the
magnitude of $U(2)$ symmetry breaking parameter $\delta m^2$.

 After the above presentation of some
intuitive arguments, we proceed with the quantitative discussion.
Let us take the conventional string configuration described in the
previous section by $\Phi_{string} = (\psi(x) = f(r) e^{i
\varphi}, \phi=0)$ and formulate the question of whether this
configuration is stable with respect to any small fluctuations. We
must stress that for $\delta m^2 >0$ topology guarantees that
small fluctuations cannot destroy the string, in the sense of
removing the windings of $\psi$ field. However, topology makes
absolutely no predictions regarding the internal structure of the
string - in particular, a priori, there is no reason for $\phi
\equiv 0$ to be a stable solution to equations of motion in the
background of the $\psi$ string.

We go along the standard procedure and expand the energy per unit
length of the system (\ref{H}) in the $\Phi_{string}$ background
to quadratic order in $\psi$ and $\phi$ modes: \beq E
(\psi=\psi_{string}+\delta\psi, ~\phi) \approx ~~\mu + \delta E,
\eeq with $\mu$ given by (\ref{mu}). We know that the $\psi$
string itself is a stable configuration so the $\delta\psi$ modes
cannot possess negative eigenvalues, which would correspond to an
instability. Therefore, we concentrate only on "dangerous modes"
related to $\phi$ fluctuations, in which case $\delta E$ is given
by:
 \begin{widetext}
\bea \label{Phi_Energy} \delta E &=& \int d^2 x (|\nabla \phi|^2 +
(f(r)^2 -1) |\phi|^2 + \delta m^2 |\phi|^2 + \frac{1}{2}
|\phi|^4) \approx \int d^2 x {\phi}^* (\hat{O} + \delta m^2) \phi
\\ \nonumber \hat{O} &=& -{\nabla}^2 + (f(r)^2 -1)\eea
\end{widetext}
where $f(r)$ is the solution of  eq.(\ref{f_Equation}), and
  we   keep only the second order terms in $\phi$
assuming that the fluctuations are small. If $\delta E$ is a
positive quantity, then the $\Phi_{string}$  is an absolutely
stable configuration and $\phi$  modes do not destroy the solution
discussed in the previous section. On the other hand, if $\delta
E$ is negative, this means that the corresponding $\phi$ is a
direction of instability in the configurational space and the
structure of the stable vortex differs from the simple
$\Phi_{string}$ solution.
  Thus, the question of whether a
negative mode exists can be answered by determining whether the
Hermitian operator $\hat{O} + \delta m^2$ possesses a negative
eigenvalue. So we have to solve the eigenvalue problem $\hat{O}
\phi = \epsilon \phi$, or more explicitly: \beq \label{Shrodinger}
[-{\nabla}^2 + (f(r)^2 -1)] \phi(x) = \epsilon \phi(x) .\eeq
Notice that this is just the 2-dimensional Schrodinger equation in
a cylindrically symmetric potential $V(r) = f(r)^2 -1$, which is
everywhere negative and tends to $0$ at infinity. It
 is well known that such a  potential
 always possesses a bound state with $\epsilon < 0$
\cite{Landau_Lifshitz}.
Therefore, if $\epsilon + \delta m^2 < 0$ then a negative mode
exists and the string obtains a non-trivial $\phi$ core. This
implies that the $\tilde{U}(1)$ symmetry gets "spontaneously
broken" in the core. The above formal fact has very important
consequences such as the existence of a Goldstone mode capable of
propagating along the core of the string. Otherwise, if $\epsilon
+ \delta m^2
> 0$, the operator $\hat{O} + \delta m^2$ is positive definite and no
core condensation occurs, in which case the conventional string
solution $\Phi_{string}$ discussed in the previous section is
valid, and both $U(1)$ symmetries are restored in the center of
the core.

We want to determine precisely the critical value $\delta m^2_c =
|\epsilon|$ below which the $\psi$ vortices obtain $\phi$ cores.
We approach this problem numerically. Since $V(x)$ is
cylindrically symmetric, we assume that the lowest energy
eigenfunction $\phi(x)$ has no angular dependence. Then the
operator $\hat{O}$ reduces to, \beq \label{O_cyllindrical} \hat{O}
= -\frac{1}{r} \frac{d}{dr} (r \frac{d}{dr}) + (f(r)^2 -1).
 \eeq
We discretize the operator $\hat{O}$ using our numerical solution
for $f(r)$ computed in the previous section. We then compute the
lowest eigenvalue of the discretized operator. This turns out to
be $\epsilon \approx -0.25$, which implies that the critical value
for $ \delta m^2$ where the transition occurs is $\delta m^2_c=
(\mu_1-\mu_2)/\mu_1 \simeq 0.25$. It is quite obvious that the
condensation of the $\phi$ field becomes even more pronounced when
$\delta m^2$ becomes smaller. Therefore, the condensation of the
$\phi$ field in the core of the $\psi$ string occurs for $0 <
\delta m^2 \lesssim 0.25$. In other words, for this range of
$\delta m^2$ the $\tilde{U}(1)$ symmetry is spontaneously broken
in the core of the string.
%
%

One should remark here that
for $\delta m^2$ close to its critical value $\delta m^2_c\simeq 0.25$,
 the condensation in the core is
small, the fourth order terms in (\ref{Phi_Energy}) can be
neglected and the eigenfunction of $\hat{O}$, which is well
localized in the vortex core, is a good approximation to
$\phi(r)$, while the term $\frac{1}{2}|\phi|^4$ in
eq.(\ref{Phi_Energy}) can be used in a variational calculation to
determine the normalization of this eigenfunction. However, when
$\delta m^2$ becomes sufficiently smaller than $\delta m_c^2$, the
$\phi$ condensation in the core becomes large, and the shape of
profile function $\phi(r)$ is influenced by the fourth order
terms.
Moreover, when $\delta m^2$ becomes very close to $0$, the
magnitudes of $\phi$ and $\psi$ fields in the core become
comparable (as expected from symmetry considerations) and the
$\phi$ field can no longer be treated as a weak perturbation to
the background $\psi$ field. In this regime,$\, \psi(x)$ and
$\phi(x)$ can be found through a full variational calculation or
by solving a coupled system of boundary value problems for $\psi$
and $\phi$ fields. We want to obtain explicit solutions for the
fields $\psi, \phi$  since they will tell us how the vortex
evolves as we approach the topologically unstable limit $\delta
m^2 \rightarrow 0$ by varying the external parameters $\mu_1,
\mu_2$. It will turn out that by answering this apparently
theoretical question we will find a way to test the
system for core condensation experimentally  by  probing
 the internal structure of the vortex.

In this paper we adopt the numerical approach to treat the above
problem. The full Euler-Lagrange equations for fields
$(\psi,\phi)$ are: \bea \label{FullPsi} {\nabla}^2 \psi &-&
(|\phi|^2 + |\psi|^2 -
1) \psi = 0   \\
 {\nabla}^2 \phi &-& (|\phi|^2 + |\psi|^2 - 1) \phi
- \delta m^2 \phi = 0 \nonumber
\eea
Assuming the simplest solution, $\psi(x) = f(r) e^{i \varphi}$,
$\phi(x) = \phi(r)$, equations (\ref{FullPsi})
reduce to:
\bea \label{FullPsiR} f'' + \frac{1}{r}f' - \frac{1}{r^2}f -
(\phi^2 + f^2 -1) f = 0 \\
\label{FullPhiR} \phi'' + \frac{1}{r} \phi' - (\phi^2 + f^2 - 1 +
\delta m^2) \phi = 0 \eea The boundary conditions on $f$ and
$\phi$ are $f(0) = 0$, $f(\infty) = 1$, $\phi'(0) = 0$,
$\phi(\infty) = 0$. The condition $\phi'(0) = 0$ is necessary for
$\nabla^2 \phi$ at the origin to be finite. Before we demonstrate
the explicit numerical solutions to equations
(\ref{FullPsiR},\ref{FullPhiR}), it is instructive to calculate
the asymptotic behavior of $\phi(r)$ at large $r$, which provides
us with a  qualitative picture of the core size.
 In this case, $1-f(r)^2$ as well as  $\phi(r)^2$ are small
so that eq. (\ref{FullPhiR}) can be  linearized to obtain: \beq
\label{LinearPhi} \phi'' + \frac{1}{r} \phi' - \delta m^2 \phi =
0. \eeq Solution of eq.(\ref{LinearPhi}) is known to be: \beq
\label{PhiAsymptote} \phi(r) \sim {\mathrm{K}}_0 (\delta m \,
r)\sim e^{-\delta m r}, \eeq where  ${\mathrm{K}}_0$ is the
modified Bessel function. The behavior of solution
(\ref{PhiAsymptote}) at large $r$
 suggests that the distance scale (in physical units), over which
$\phi$ condenses is of order $\xi/{\delta m}$ rather than the
correlation length $\xi$ . Hence we expect that as $\delta m^2
\rightarrow 0$ the width of $\phi$ condensate in the core grows.
But $\phi$ and $\psi$ particles
 always try to minimize their
common territory. Thus, we expect that the region of the string
where $|\psi| \neq 1$ also has the length scale $\xi/{\delta m}$
rather than the typical correlation length $\xi$. As we will
demonstrate below, our numerical results confirm this intuitive
reasoning. We conclude that as $\delta m^2 \rightarrow 0$, the
core of the string expands and eventually fills the whole space.
So the string disappears in the limit $\delta m^2 = 0$ in
accordance with topological arguments.

\begin{figure}
\label{Fig1}
\epsfxsize=80mm
\epsfbox[ 58   200   548   589]{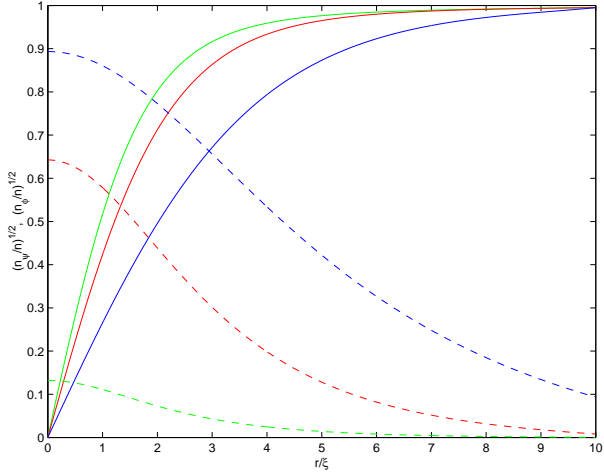}

\caption{Vortex profile function $|\psi(r)|$ (solid) and core
condensate profile function $\phi(r)$ (dashed) for different
values of asymmetry parameter $\delta m^2 = (\mu_1 -
\mu_2)/\mu_1$. Green, red and blue curves correspond to $\delta
m^2 = 0.25, 0.15$ and $0.05$ respectively. The quantities plotted
are square roots of particle densities (in units of bulk density
$n$) as a function of distance from the vortex core center $r$ (in
units of correlation length $\xi$).}

\end{figure}

Let us now explicitly test the above predictions. We solve
numerically the system  (\ref{FullPsiR},\ref{FullPhiR}) for
different values of $\delta m^2$. Fig. 1 displays the numerical
solutions $f(r)$ and $\phi(r)$ for $\delta m^2$ = $0.25, 0.15$ and
$0.05$. Essentially, Fig. 1 demonstrates how the densities of
condensed particles
 $ \sqrt{n_{\psi}(r)/n}=|\psi(r)|$
and   $ \sqrt{n_{\phi}(r)/n}=|\phi(r)|$ vary near the vortex core
with distance measured in units of the correlation length $\xi$.
From Fig. 1 it is evident that as $\delta m^2$ decreases, the
magnitude and width of $\phi$ condensation in the core increases
considerably, and the core of the $\psi$ vortex expands. Thus, the
numerical calculations support our intuitive explanation given
above. The expansion of the vortex core due to core condensation
of the second component may have very important practical
consequences. Typically, the core size in BEC experiments cannot
be directly observed because $\xi \leq 1 \mu m$, which usually is
below the resolution limit. We speculate at this point that one
can adjust the asymmetry of the system in such a way that the core
condensation occurs. Further decreasing  the asymmetry by
adjusting $\mu_1, \mu_2$, one can  make the core large enough
(larger than the resolution limit) to allow direct detailed
experimental measurements of vortex core properties, such as the
core radius and the density of the second component $\phi$ filling
the core.

Summarizing the results of this section, we have demonstrated that
in the case when the $U(2)$ symmetry of Lagrangian
(\ref{Lagrangian}) is broken only weakly, the $\psi$ strings might
obtain non-trivial cores. Inside of these cores, the $\phi$ field
is condensed and the $\tilde{U}(1)$ symmetry is spontaneously
broken. We have found that the necessary and sufficient condition
for core condensation is $0 < \delta m^2 \lesssim 0.25$, which
 implies that the effect takes place even if the
difference in chemical potentials of the two components is as
large as $25\%$. The described effect becomes much more pronounced
if the asymmetry parameter is small. When the asymmetry  $\delta
m^2 \rightarrow 0$, the string core expands and eventually fills
the whole space, destroying the topological defect.


\section{Vortons}
\subsection{Stable String Loops}
The most important result of the previous section is that under
certain conditions a
 $\phi$ condensate will be trapped inside the $\psi$ string. In this case,
according to Goldstone's theorem, a new degree of freedom
corresponding to the phase of the $\phi$ field becomes a massless
  Goldstone boson, which can travel along the core of the string with arbitrarily
low energy cost.

Now, we follow Witten \cite{Witten}, who has constructed his
superconducting string by making a loop out of the vortex with a
second condensate in the core. Therefore, let us   consider
bending the $\psi$ string with a $\phi$ core into a loop of large
radius $R$. In order to ignore curvature effects, we take $R$ to
be much greater than the width of the string core. Such a
configuration is certainly unstable from the topological point of
view since the change of argument of $\psi$ along a large circle
surrounding the loop is always $0$ (so the loop belongs to the
trivial topological class). From another point of view, the energy
of the loop is $E = 2 \pi \mu R$, where $\mu$ is the string
tension computed in Section II. Thus, the energy $E$ is minimized
when $R = 0$. The fact that a simple string loop shrinks with time
in the relativistic setting has been confirmed by numerical
simulations\cite{DavisShell4}, and the same behavior should be
expected in the non-relativistic case provided the loop does not
move and the Magnus force does not equilibrate the string tension.

Nevertheless, as was originally proposed for the case of cosmic
strings \cite{Hindmarsh,Turner,Davis, DavisShell1, DavisShell2,
DavisShell3, Shellard2002}, a string loop can be stabilized by a
persistent current carried by the phase of the core condensate
$\phi$ along the loop. Here we discuss this original idea applied
to our non-relativistic system. Indeed, suppose that the phase of
the $\phi$ field varies by $2 \pi N$, as one traverses the string
loop, where $N$ is constrained to be an integer since $\phi$ is
single valued. Then $N$ acts as a conserved number: in order for
$N$ to change, the field $\phi$ has to vanish somewhere on the
string core, which is energetically unfavorable by the results of
the previous section. $N$ is said to be a semi-topological charge
since it is conserved in part due to topological reasons and in
part due to energetics. We can explicitly parameterize the field
$\phi$ of this configuration as $\phi (x) = \phi (r) e^{i k z}$,
where $r$ is the distance to the core center in the plane
perpendicular to the string direction, so that $\phi(r)$ is the
same as in the previous section, and $z$ is the spatial variable
running along the string core. Then the variation of the phase of
$\phi$ along the string loop is $2 \pi k R$, so we can identify,
$k = \frac{N}{R}$.

In a non-relativistic system, there is always present another
conserved number  - the Noether's charge associated with the
$\tilde{U}(1)$ symmetry, which is simply the total number of
$\phi$ particles. This is given by, \beq \label{Charge} Q = \int
d^3 x |\phi|^2 \eeq
   As we
demonstrated previously, it is unfavorable for $\phi$ particles to leave
 the string, so the conserved charge $Q$ will be
confined to the string core, being effectively trapped by the dynamics
rather than by some external forces.

Notice that the variation of the phase of $\phi$ along the string
influences the energy of the configuration through the term
$|{\d}_z \phi|^2$ in the energy density. When integrated over
3-dimensional space, this term contributes the following amount
into the energy of the system, \beq \label{dz_term} \Delta E = k^2
\int d^3 x |\phi|^2 = \frac{N^2}{R^2} Q. \eeq
 Thus, ignoring other contributions for a moment (detailed analysis
of this problem will be presented in section IV C),
 the energy of the loop becomes,
\beq \label{ApproxE} E \approx 2 \pi \mu R + \frac{N^2}{R^2} Q.
\eeq
Thus, as the charges $N$ and $Q$ are conserved within the configuration,
 the energy of
the vortex loop has a non-trivial minimum with respect to the loop
radius at $R_0 = \sqrt[3]{N^2 Q/\pi \mu}$. Therefore, it is
expected that the loop will stop shrinking at the above radius and
a classically stable configuration known as vorton will form. The
stability is purely mechanical in origin and can be thought to be
due to the angular momentum $M = Q N$ carried by the vorton (see
section IV D).

The formula (\ref{ApproxE}) is quite important, so we want to
restore all dimensional factors to represent the final result as
follows, \bea \label{Edimensional} \tilde{E}(\tilde{R_0})=
3\pi\tilde{\mu} \tilde{R_0},~~~~ \tilde{R_0}=
\sqrt[3]{\frac{\hbar^2}{2m}\frac{N^2 \tilde{Q}}{\pi \tilde{\mu}}}
\eea where $\tilde{\mu}$ is determined by   expression
({\ref{mupsi}) and $\tilde{Q}$ is the physical number of trapped
$\phi$ particles. The most striking feature of the vorton is that
its size could, in principle, be arbitrarily large. Practically,
the size is mainly limited by the boundary conditions specific to
a given experiment and by our ability to produce a configuration
with a large winding number $N$ and charge $Q$ during the
formation period.

It is important to note that the vortons described here have
nothing in common with vortex loops (observed in the superfluid
$^4He$, see \cite{Rayfield})
 stabilized by Magnus force, which is
the interaction of a moving string with its environment. String
rings discussed in the present work
 can be stable even if they do not move as a whole object.
The non-relativistic configuration bearing some similarities to
the one described here is the twisted vortex ring (vorton)
discussed in the superfluid $^3He - B$ system at low temperature
and pressure, see \cite{VolovikBook} and references therein.
Although it was originally hoped that vortons can become stable in
this system, numerical estimates have shown\cite{VolovikBook}
 that it is unlikely to happen,
and such rings have never been observed in $^3He - B$.
Actually, vortex ring formation and dynamics in BECs were
discussed earlier\cite{Jackson1, Jackson2, Feder, Komineas1,
Komineas2}. However, in most cases the analysis of ring
stabilization (as opposed to formation) was done for a
one-component BEC, where the stability (or meta-stability) is due
to the combination of ring motion and details of trapping
potential. In contrast, our main objective is the analysis of
vortons, which can be stable in an infinitely large volume without
moving; this is the kind of system, which might be of interest for
cosmology and astrophysics.
 A close analogue of a vorton is the
full 3D skyrmion, which was recently discussed in the context of
2-component BECs\cite{Anglin,Battye,Stoof,Ruostekoski}. Similarly
to a vorton, the skyrmion can be represented as a vortex ring in
one component with the second component carrying an integer quanta
of circulation around the ring.
Numerical simulations\cite{Battye,Ruostekoski} of the skyrmion
solution in BECs resemble very much the analytical construction of
vortons discussed here. We suspect that the  stability of
configurations considered in \cite{Battye,Ruostekoski} is  due to
the energetics rather than topology, which is claimed to be the
main source of stability,
 see comment\cite{skyrmion}. We must stress that these simulations were
performed for small values of topological charge only, thus, producing
point-like objects of small size. On the other hand, the
size of vortons is determined by quantum numbers $Q$ and $N$ and
could, in principle, be arbitrarily large as explained below.

\subsection{Relativistic Springs and Vortons}

The detailed construction of vortons in non-relativistic physics
will be discussed in the next subsection. However, we believe that
a historical aside will be instructive here: we now make a detor
to briefly describe stable vortex loops, which were first
considered in the context of cosmic strings
\cite{Hindmarsh,Turner} (for a full review of the subject see
\cite{Shellard2002}). Later on, the ideas developed in the above
works have been applied to systems with high baryon
density\cite{KVortons} and to high $T_c$
superconductors\cite{SO5Vortons}. The lessons learned from these
papers prove to be very useful for our discussion of vortons in
non-relativistic systems.

In this subsection, we consider the phenomenon of string core
condensation and vorton stability in the
 relativistic system
defined by the following Lagrangian, \beq \label{RelLagr} {\cal L}
= |\d_0 \Phi|^2 - { \cal H }, \eeq where ${\cal H}$ is the same as
before (\ref{H}). Here we use the relativistic notation with
$\hbar=c=1$. The Noether's charge associated with the
$\tilde{U}(1)$ symmetry in the relativistic case is: \beq
\label{RelCharge} Q = \frac{i}{2} \int d^3 x (\d_0 \phi \,
{\phi}^* - {\d_0 \phi}^* \, \phi). \eeq It is important to note
that  in the relativistic setting, the $\phi$ condensate in the
core of the string does not necessarily carry a $\tilde{U}(1)$
Noether's charge. In particular, all static solutions satisfy $Q =
0$.

The first attempts to stabilize the string loops were based on the
introduction of a winding number density $k = N/R$ of the core
condensate $\phi$, exactly as described in the previous section.
These relativistic objects, dubbed springs, were characterized by
$N \neq 0$ and $Q = 0$. It was soon realized, however, that such
objects, in general, are not stable: persistent current related to
the winding number $N$ cannot stabilize the string loops against
shrinking. Fortunately, a related mechanism which makes string
rings stable with respect to classical decay was found  by Davis
and Shellard \cite{Davis, DavisShell1, DavisShell2, DavisShell3}.
The idea is to provide the configuration with a nonzero charge
(\ref{RelCharge}) trapped in the string core along with the
winding number $N$; such a configuration carries an angular
momentum $Q N$, hence its name - "vorton". According to
(\ref{RelCharge}), the presence of charge $Q$ automatically
implies time dependence of the $\phi$ field, which can be
parameterized as $\phi(\vec{x},t) = e^{i (k z - \omega t)}
\phi(r)$, where $\phi(r)$ is the profile function similar to one
discussed previously. Naively, one could think that time
dependence of a classical solution brings in additional energy
into the system, which usually further undermines stability.
However, as one can show, the stability is enforced by
conservation of charge $Q$. In a sense, the time-dependent
configuration becomes the lowest energy state in the sector with
given quantum numbers: non-zero charge $Q$ and winding number $N$.
A similar time-dependent ansatz for a different problem was
discussed by Coleman in \cite{qball} where he introduced the
so-called Q-balls: macroscopically large stable objects with time
dependent fields.
It is currently believed that similarly to Q-balls, relativistic
vortons are classically stable objects. The recent, full scale
numerical simulations\cite{Shellard_num} support this statement.
In the non-relativistic models to which our paper is devoted, a
Noether's charge $Q$ is always present when the field $\phi$ is
condensed in the core, so the vorton stabilization mechanism is
very similar to the one suggested by Davis and Shellard.

\subsection{Non-relativistic Vortons}
Let us now demonstrate in detail how the string loop shrinks and
how the vorton stability is achieved in a non-relativistic system.
We are interested in configurations characterized by two quantum
numbers, \beq \label{NonRelCharges} N = \frac{1}{2 \pi} {\int}_{C}
d \arg \phi(x), \quad Q = \int d^3 x |\phi|^2 \eeq where $C$ is
the path along the core of the string loop. $N$ is the topological
charge, and $Q$ is a $\tilde{U}(1)$ Noether's charge, which is
just the total number of type $\phi$ particles. Fixing the charges
$Q$, $N$, and also for now fixing the loop radius $R$, we wish to
determine the corresponding fields. We can adopt an ansatz for the
field $\phi$ motivated previously, \beq \label{VortonAnsatz}
\phi(x,t) = e^{i (k z - \omega t)} \phi(r) \eeq where $\phi(r)$ is
the radial profile of the core condensate in the plane
perpendicular to the string direction. As already explained, $k$
acts as a winding number density, and $k$ = $N/R$. As we will see
shortly, $\omega$ can be identified with the Lagrange multiplier
for conservation of charge $Q$ within the configuration under
discussion. At this point, $\omega$ is still a free parameter; it
will be later eliminated by solving the equations of motion using
the ansatz (\ref{VortonAnsatz}) and substituting the solution into
the constraint for conservation of charge $Q$
(\ref{NonRelCharges}). A more precise physical meaning of $\omega$
will be given later in the text.

Substituting the ansatz (\ref{VortonAnsatz}) into the action
corresponding to Lagrangian (\ref{Lagrangian}), we obtain: \beq
\label{Action} S[\Phi] = \int dt ( \omega \int d^3 x |\phi(x)|^2 -
\int d^3 x {\cal H}[\Phi]) \eeq where ${\cal H}$ is the full
Hamiltonian density of the system defined by eq. (\ref{H}). The
Lagrange multiplier nature of $\omega$ is now evident - the energy
of the system is minimized subject to constraint $\int d^3 x
|\phi(x)|^2= Q$.
 In terms of the core condensate profile $\phi(r)$ (see eq. \ref{VortonAnsatz}),
  the above constraint requires, \beq
 \label{Sigma2Const} \Sigma_2 \equiv \int d^2 x |\phi(r)|^2 =
\frac{Q}{2 \pi R} \eeq The physical meaning of the parameter
$\Sigma_2$ introduced above is quite obvious: it is the number of
$\phi$ particles trapped in the core per unit length of the
string. In physical units, $$\widetilde{\Sigma}_2=
\frac{\tilde{Q}}{2\pi\tilde{R}} = \int d^2 \tilde{x}
|\tilde{\phi}|^2 = n\xi^2\Sigma_2$$ where $\tilde{Q}$ is the
physical total number of trapped $\phi$ particles and $\Sigma_2$
is the dimensionless number in eq. (\ref{Sigma2Const}).

We want to construct an effective Lagrangian describing the
"dynamics" of the field $\phi(r)$ on the two-dimensional $xy$
plane perpendicular to the string axis, which points in the $z$
direction, see (\ref{VortonAnsatz}). The effective Lagrangian is
obtained by substituting the ansatz (\ref{VortonAnsatz}) into the
Lagrangian (\ref{Lagrangian}), \bea \label{L_eff} {\cal L}_{eff}
&=& -|\nabla \Phi|^2 - \frac{1}{2}(|\Phi|^2
- 1)^2 -  {\widehat{\delta m}}^2 |\phi|^2 ,\nonumber \\
{\widehat{\delta m}}^2 &\equiv& (\delta m^2 + k^2 - \omega),
\eea where the gradient is with respect to two coordinates $(x,y)$
perpendicular to the axis of the string. Hence the dynamics of the
system on the $xy$ plane are determined by the new effective
parameter $ {\widehat{\delta m}}^2$
   defined as follows:
\beq \label{deltameff} {\widehat{\delta m}}^2 = ((\mu_1-\mu_2) +
\frac{\hbar^2}{2m}\tilde{k}^2 - \hbar\tilde{\omega})/\mu_1 , \eeq
where we restored in this expression all   dimensional factors to
stress its physical significance. This parameter plays the same
role as the $U(2)$ asymmetry ${\delta m}^2$ in section III: the
equations of motion for profile functions $\psi(r)$ and $\phi(r)$
imposed by the Lagrangian (\ref{L_eff}) are the same as eq.
(\ref{FullPsi}), but with $\delta m^2$ replaced by
$\widehat{\delta m}^2$: \bea \label{FullPsiV} {\nabla}^2 \psi &-&
(|\phi|^2 + |\psi|^2 -
1) \psi = 0   \\
\label{FullPhiV}
 {\nabla}^2 \phi &-& (|\phi|^2 + |\psi|^2 - 1) \phi
- \widehat{\delta m}^2 \phi = 0  \quad \eea where all derivatives
are with respect to coordinates $(x,y)$ in the plane perpendicular
to string axis.

\begin{figure}
\begin{center}
\label{Fig2}
\epsfxsize=80mm
\epsfbox[ 55   198   543   583]{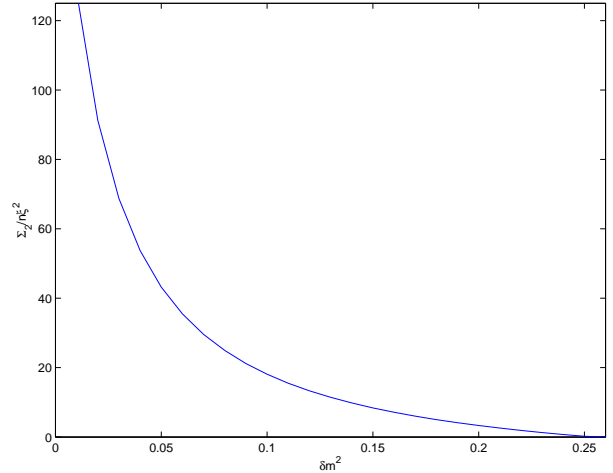}
\end{center}
\caption{Number of $\phi$ particles trapped in the vortex core per
per unit vortex length, ${\Sigma}_2$ (measured in units of
$n\xi^2$) as a function of effective asymmetry parameter
${\widehat{\delta m}}^2 = (\mu_1 - \mu_2 + \frac{\hbar^2 k^2}{2 m}
- \hbar \omega)/\mu_1$.}
\end{figure}

Following the procedure of section III, for each value of
$\widehat{\delta m}^2$, we numerically solve the equations of
motion (\ref{FullPsiV},\ref{FullPhiV}) subject to string boundary
conditions, to determine the profile functions $\psi(r)$ and
$\phi(r)$. We then use the solution $\phi(r)$ to find $\Sigma_2$
(see eq. (\ref{Sigma2Const})) as a function of $\widehat{\delta
m}^2$. Fig. 2 shows how the $\phi$ particle density per unit
vortex length, $\Sigma_2 = \widetilde{\Sigma}_2/(n\xi^2)$, depends
on the effective asymmetry parameter $\widehat{\delta m}^2$. As
one can see from this plot, the magnitude of $\Sigma_2$ increases
drastically when $\widehat{\delta m}^2$ decreases. This is due to
the combination of two effects. First, the magnitude of the $\phi$
field in the core increases when $\widehat{\delta m}^2$ decreases,
as shown in Fig. 1. Second, the core size itself increases with
decrease in $\widehat{\delta m}^2$. Also, as Fig. 2 demonstrates,
$\Sigma_2$ vanishes when $\widehat{\delta m}^2 > \delta m_c^2
\approx 0.25$, since as was demonstrated in section III, the
condensation of the $\phi$ field in the core becomes unfavorable
for large values of $U(2)$ asymmetry parameter.

Now we wish to understand the physical meaning of the parameter
$\omega$, which was originally introduced in a formal way as a
Lagrange multiplier. We define our configuration by fixing the
quantum numbers $Q$, $N$ and the loop radius $R$. To possess these
quantum numbers the fields must satisfy the constraints
(\ref{NonRelCharges}), which fix the values of $k = N/R$ and
$\Sigma_2 = Q/2 \pi R$. Yet, as we have shown above, there is a
one-to-one correspondence between $\Sigma_2$ and the effective
symmetry breaking parameter $\widehat{\delta m^2}$ (see Fig. 2).
Thus, the value of $\widehat{\delta m^2}$ is also fixed by the
external parameters $(Q,N,R)$. This, in turn, implies that
$\omega$ is fixed via eq. (\ref{L_eff}) and hence $\omega$ is
completely determined by initial assignments of $(Q,N,R)$. Now
trading external parameters $(Q,N,R)$ for $(Q,N,k)$, we speculate
that $\omega(k)$, implicitly determined by eq. (\ref{L_eff}), is
just a complicated dispersion relation describing how the
Goldstone mode propagates along the string loop in the presence of
charge $Q$ and winding number $N$.

Our next task is the analysis of energetics of the vortex loop,
with the goal of finding how the ring energy depends on its radius
$R$. In what follows we apply the technique developed
in\cite{Shellard2002} to the analysis of our non-relativistic
system. Starting from the Lagrangian (\ref{Lagrangian}), we switch
to a Hamiltonian formalism and represent the total energy
 of the configuration in the following way,
\begin{widetext}
\bea
\label{VortonphiE} E  &=& 2\pi\mu R+\int d^3 x (|\nabla \phi|^2 +
(|\psi|^2 -1 + \delta m^2) |\phi|^2 + \frac{1}{2}|\phi|^4) \\
\nonumber &=& 2\pi\mu R +2 \pi R \int d^2 x [{\phi}^* (-{\nabla}^2 + |\psi|^2
- 1 + \delta m^2 + k^2) \phi + \frac{1}{2}|\phi|^4], \eea
\end{widetext}
where $2\pi\mu R$ term is the conventional contribution
(\ref{mupsi}) due to the $\psi$ field, and in the last step we
have performed an integration by parts. The laplacian in the last
equation is taken with respect to the $(x, y)$ variables
perpendicular to the string axis. The above expression can be
simplified via the Euler-Lagrange equation for the field $\phi(r)$
(\ref{FullPhiV}). Multiplying (\ref{FullPhiV}) by the complex
conjugate $\phi^*(r)$ and then inserting the result into
(\ref{VortonphiE}), we find: \bea \label{VortonPhiES} E = 2\pi\mu
R+ 2 \pi R \int d^2 x \Big(\omega
|\phi|^2 - \frac{1}{2} |\phi|^4\Big)  \nonumber \\
=2\pi\mu R+ 2 \pi R \Big(\omega \Sigma_2 - \frac{1}{2}
\Sigma_4\Big), ~~~~~~~~~~~~~~ \eea where $\Sigma_4\equiv \int
d^2x|\phi|^4$. A similar expression for the relativistic vortons
was derived in\cite{Shellard2002}. As we discussed earlier,
$\omega$ in this expression can be computed by inverting the
constraint (\ref{Sigma2Const},\ref{L_eff}) producing a complicated implicit
function of $k$, which is difficult to analyze in general.
However, for relatively large $\omega$, such that $\omega
> \delta m^2, \widehat{\delta m^2}$ (which are smaller than $\delta m_c^2 \approx 0.25$)
one can use eq. (\ref{L_eff}) to approximate $\omega \approx k^2 =
N^2/R^2$. The physical meaning of this simplification is that the
typical energy scale of the $\phi$ field excitations along the
string is larger than the typical scale of excitations
perpendicular to the string. To further simplify our discussion,
we would like to consider the simple case when the core
condensation is not very large and, therefore, the back reaction
of the $\phi$ field onto the $\psi$ vortex can be neglected, so
that one can use the unperturbed expression (\ref{Var_Soln}) for
the string tension $\mu$ (\ref{mupsi}).  Also, one can check that
in this regime $\Sigma_4 \ll \Sigma_2$, so the last term in
equation (\ref{VortonPhiES}) can be dropped, and
  the total energy of the configuration becomes,
\beq \label{EVortonF} E = 2 \pi \mu R + \frac{N^2}{R^2} Q,
 \eeq
where we replaced $\omega \rightarrow  k^2 = N^2/R^2$. It is
interesting to note that the equation (\ref{EVortonF}) exactly
coincides in form with the one (\ref{ApproxE}) derived with a
number of simplifications mainly for demonstrative purposes. Thus,
as was originally pointed out, conservation of charges $Q$ and $N$
guarantees that the loop energy $E$ has a non-trivial minimum with
respect to radius $R$ at,
\bea \label{R0} R_0 = \sqrt[3]{\frac{N^2 Q}{\pi \mu}}, \quad
E(R_0) = 3 \pi \mu R_0 \eea The corresponding expressions with
restored dimensional parameters were presented earlier in
(\ref{Edimensional}). As we argued above, we expect that at this
radius the string loop will become stabilized against further
shrinking, forming the so-called "vorton". The winding number
density $k$ and the particle number density $\Sigma_2$
corresponding to the
  radius $R_0$ are:
\bea
\label{k0} k = {\sqrt[3]{ \pi \mu}}\cdot
\sqrt[3]{\frac{N}{Q}}, \nonumber \\
\quad \Sigma_2 =  \frac{\sqrt[3]{ \pi \mu}}{2\pi}\cdot
 \sqrt[3]{\Big(\frac{Q}{N}\Big)^2}.
\eea Due to the importance of these expressions as explicitly
measurable quantities, we would like to present them with all
dimensional factors included, \bea \label{kdimensional} \tilde{k}
= \sqrt[3]{\frac{2m\pi \tilde{\mu}}{\hbar^2}}\cdot
\sqrt[3]{\frac{N}{\tilde{Q}}}, \nonumber \\
\quad \widetilde{\Sigma}_2 =  \frac{1}{2\pi}\sqrt[3]{\frac{2m\pi
\tilde{\mu}}{\hbar^2}}\cdot
 \sqrt[3]{\Big(\frac{\tilde{Q}}{N}\Big)^2}.
\eea One immediate observation is that if one neglects a somewhat
weak $\ln R$ variation of $\mu$ with $R$,
  see (\ref{Var_Soln}),
 the local parameters such as
$k$ and $\Sigma_2$ depend solely on the ratio $Q/N$. Hence, the
local parameters are approximately scale-invariant under the
transformation $(Q,N) \rightarrow (\alpha Q, \alpha N)$, where
$\alpha$ is an arbitrary transformation parameter. Conversely, the
global characteristics such as the radius $R_0$ and the energy $E$
 are extensive parameters and transform as $(R,E) \rightarrow
(\alpha R, \alpha E)$.  Thus, one should expect the existence of
stable vortons of arbitrarily large size. This statement, of
course, should be corrected when finite volume effects, such as
the trapping potential, are taken into consideration.

One can present  the arguments showing that
the back-reaction of
the $\phi$ field onto the $\psi$ vortex is negligible
for vortons in
a large region of $(Q,N)$ parameter space, where, in particular,
 the parameter $\omega$ is large compared to $\delta m^2,
\widehat{\delta m^2}$. Thus, both assumptions made in our
discussion of energetics of vortex loops with core condensates are
justified a-posteriori. One should note that when either of these
two assumptions is not valid, we still expect that stable vortons
will generally exist. However, the analysis of vorton stability
becomes very involved in this case as the string tension $\mu$ and
the frequency $\omega$ become very complicated functions of
$(Q,N,R)$, and no analytical control is possible. We have not
discussed this case because our main goal was to demonstrate the
fact of vorton existence in a large region of the parametrical
space rather than a complete analysis of the allowed region of
parametrical space where vortons are stable.
\subsection{Vortons and Angular Momentum}
In this section we demonstrate that the non-relativistic vorton
carries an angular momentum.
From Lagrangian (\ref{Lagrangian}), the total angular momentum of
the system is: \beq \label{AngularM} \vec{M} = \int d^3 x
\Big({\Phi}^{\dagger} \, [\vec{x} \times -i \vec{\nabla
}]\Phi\Big). \eeq When the $\Phi$ field forms a vorton, there is
no net circulation of $\psi$ particles and the only contribution
to the angular momentum comes from the $\phi$ field condensed in
the core.
 We can re-parameterize the
ansatz (\ref{VortonAnsatz}) as: \beq \label{AngularAnsatz} \phi(x)
= e^{i N \theta} \phi (\rho,z) \eeq where $(\rho, \theta, z)$ are
the polar coordinates with respect to the center of the vortex
loop. Substituting this into (\ref{AngularM}) gives: \beq
\label{phiM} \vec{M} = N \hat{z} \int d^3 x |\phi|^2 = Q N \hat{z}
, \eeq which is a very simple relation that can be easily
understood on the intuitive level in terms of classical physics.

Therefore, the vorton carries an angular momentum perpendicular to
the string loop plane. The magnitude of the angular momentum is
proportional to classically conserved charges $Q$ and $N$ . Thus,
as originally suggested in the relativistic context by
\cite{DavisShell2}, the stability of vortons can be understood in
terms of conservation of angular momentum.
Let us note that due to the scaling properties of eqs.
(\ref{EVortonF}, \ref{phiM}) it is energetically favorable for two
identical vortons, each with an angular momentum $M$, to merge
into one vorton with an angular momentum $2 M$, thereby increasing
the vorton radius (this statement is subject to the assumption
that that the number of $\phi$ particles trapped in the ring core
is conserved in this merge). So we expect that a typical vorton
will possess, in general, the largest possible size constrained
only by the boundary conditions and/or by the properties of the
trapping potential.

\section{Conclusion}

The main goal of this work is to demonstrate that in a
non-relativistic system, such as a two component BEC, with a $U(1)
\times \tilde{U}(1)$ symmetry and an approximate $U(2)$ symmetry,
 stable strings with a non-trivial core may exist. Inside this
core the $\tilde{U}(1)$ symmetry is spontaneously broken allowing
a Goldstone boson to travel along the string axis. We have
explicitly demonstrated that within our model the condensation
inside the core occurs even for relatively large values of $U(2)$
symmetry breaking parameter, $\delta m^2 \lesssim 0.25$.  The
effect described becomes much more pronounced when the $U(2)$
asymmetry becomes even smaller. We should note that the phenomenon
when a string core possesses a condensate of a different
field/phase is by no means a completely new situation in physics.
A similar phenomenon takes place in $^3 He$ and in high $T_c$
superconductors where it has been studied experimentally quite
extensively. Nevertheless, it would be very interesting to study
this very nontrivial phenomenon experimentally  in two component
BEC systems, with the specific focus on the dependence of the
effect on the asymmetry
 parameter $\delta m^2$. In particular, we believe that with
 decrease in asymmetry, the vortex core will expand
above the resolution constraints, therefore, providing an
 opportunity to experimentally study the core in detail.

In the second part of the paper we applied the idea of
$\tilde{U}(1)$ symmetry breaking inside the core to construct
stable string loops known as vortons. These vortons are stabilized
by a winding number carried by the core condensate field around
the loop as well as a Noether's charge associated with the
$\tilde{U}(1)$ symmetry trapped in the string core. We have seen
that the energy of such a string loop will have a non-trivial
minimum with respect to the loop radius where the loop will become
stabilized against further shrinking. We have also shown that
non-relativistic vortons carry an angular momentum, which can be
seen as the source of their classical stability.

It is puzzling that although, in general, the vorton is expected
to be a fairly common configuration in quantum field theory, as
far as we can tell, it has never been observed in nature. Indeed,
at this point, discussions of cosmic strings\cite{Witten} in
cosmology and vortons\cite{KVortons} in high baryon density
systems should be considered as no more than beautiful
speculations. Some systems proposed earlier to test the existence
of vortons are high $T_c$ superconductors\cite{SO5Vortons} and
liquid $^3 He$. However, currently no direct observations can be
presented to support the existence of vortons in high-$T_c$
superconductors, while in the $^3 He$ system, estimates suggest
that vortons cannot be stabilized \cite{VolovikBook,Volovik}.
Thus, the BEC system might provide us with the first true
experimental opportunity to prove the existence and stability of
vorton configurations that have been the subject of intense
theoretical discussions for the past 20 years.

It is tempting to identify the vortex rings observed in the BEC
system\cite{ring} with vortons, which are the subject of this
work. However, such an identification would be premature.
Actually, in the original experiment, a two-component system was
initially constructed, but the second component was removed before
the rings were created. Hence the experimental results were
compared by \cite{ring} to a simulation of a one component
BEC\cite{Feder}, where vortons cannot exist, as there is no second
component capable of condensing in the vortex core. If, indeed,
the observed rings are due to one field component only then they
may be analogous to the vortex rings in $^4 He$\cite{Rayfield},
where the ring stability is due to the Magnus force. In this case,
the rings must move in order to equalize the string tension that
causes them to shrink. Ring stability might also be due in part to
the artificial trapping potential\cite{Komineas1, Komineas2}; this
situation is not addressed in our work.

However, granted that some population of the second component
remained in the original experiment, the vortex rings observed in
\cite{ring} could be also real vortons, which are stable due to
the condensation of the second field in the core. In this case it
would be the first ever observation in nature of such objects.
There is a "smoking gun" signature, which distinguishes between
conventional vortex rings, similar to the ones observed in $^4
He$, and vortons, which have never been previously observed. This
signature is the angular momentum (\ref{phiM}), which is always
present for vortons and absent for conventional vortex
rings\footnote{The angular momentum of a conventional ring is due
to the contribution of the $\psi$ field into $M$ in
eq.(\ref{AngularM}), which is identically zero.}. We do not know
whether the angular momentum of the ring (or equivalently the
circulation of the second component in the core of the ring) can
be measured in experiments similar to \cite{ring}. Another vorton
signature is its ability to stay still for a relatively long
period of time. This distinguishes vortons from the conventional
vortex rings, which shrink and decay in the absence of motion. We
do not know whether the rings, which typically move when they are
formed, can be stopped to observe their behavior. However, if
either of the two experiments described above can be performed, it
would be the "smoking gun" evidence demonstrating the existence of
vortons stabilized due to core condensation.


Finally, if vortons can, indeed, be formed and studied in BEC
experiments, it would provide a unique opportunity to study not
only condensed matter physics, atomic physics and optics,
 but also astrophysics and cosmology
by doing laboratory experiments. In particular, one would finally
be able to test mathematical ideas pioneered by Witten, Turner,
Kibble, Davis, Shellard and many others. Over the last few years
several experiments have been done to test ideas drawn from
cosmology (see the review papers
\cite{kibble,volovik,VolovikBook,Volovik} for further details).
One can further speculate on a possibility to study the formation,
evolution and stability of different topological objects
(including vortons) relevant for cosmology
 by doing controlled experiments with BEC
systems as discussed in the recent workshop\cite{COSLAB}.

\section*{Acknowledgments}
We are thankful to Kirk Madison for his talk on vortices in BECs
given at UBC, which strongly motivated this study. We are also
grateful to him for sharing with us his vision of many
experiments, which have already been performed (or can be done in
the nearest future) in the laboratory. We are also thankful to
Grisha Volovik, Janne Ruostekoski, Shou-Cheng Zhang and other
participants of the COSLAB workshop\cite{COSLAB} for their
insights and useful discussions. This work was supported in part
by the Natural Sciences and Engineering Research Council of Canada
and the ESF Program "Cosmology in the Laboratory" (COSLAB).

\appendix
\section*{ Appendix: Physical Scales}
In this Appendix, we reduce the Lagrangian (\ref{DimLagMod}) to a
dimensionless form (\ref{Lagrangian}). In order to do this, we
introduce scaled coordinates and fields,
\begin{widetext}
\bea \label{scale} \tilde{t} = \frac{t}{\omega_0}, ~
\vec{\tilde{x}} = \xi \vec{x}, ~ \tilde{\Phi} = \sqrt{n} \Phi, ~
\omega_0 = \frac{\mu_1}{\hbar}, \xi = {\Big(\frac{\hbar^2}{2 m
\mu_1}\Big)}^{1/2},~n = \frac{\mu_1}{g}, ~\delta m^2 = \frac{\mu_1
- \mu_2}{\mu_1} \eea
\end{widetext}
We remind the reader that the original (dimensionfull) variables
carry a "tilde", while the introduced dimensionless variables
carry no "tilde". In the above notation, the action $\tilde{S}$
takes the form,
\begin{widetext} \bea \label{action}
  \frac{\tilde{S}}{\hbar} =(n\xi^3) \int dt d^3x (i
\Phi^{\dagger} \d_0 \Phi - |\nabla {\Phi}|^2 -
\frac{1}{2}(|\Phi|^2 -1)^2 - \delta m^2 |\phi|^2) ~~~~~~~~~~
\eea
\end{widetext}
where all derivatives are with respect to dimensionless
coordinates $(t, \vec{x})$. This is exactly (up to a
multiplicative factor) the action corresponding to the
dimensionless Lagrangian (\ref{Lagrangian}) used throughout this
paper. The physical distance corresponding to one unit of
dimensionless variable $x$ is just the correlation length $\xi$,
the physical time corresponding to one unit of dimensionless
variable $t$ is $1/\omega_0$ and the physical density
corresponding to one unit of $|\Phi|^2$ is just the bulk BEC
density $n$. The typical values of parameters in experiments on
$^{87}Rb$ are $\xi \sim 0.1 \mu m$, $n \sim 10^{14} cm^{-3}$,
$\omega_0 \sim 5 kHz$


\begin{thebibliography}{200}
\bibitem{skyrmion} Indeed, the stability
of configurations in refs.\cite{Anglin,Battye,Ruostekoski} is due
to the asymmetry of the interaction potential expressed by the
parameter $\Delta \neq 0$ in \cite{Battye}. It is quite obvious
that if the
  stability of a configuration
is based on a nonzero value of asymmetry parameter $\Delta$,
  the topological arguments are irrelevant in the explanation
of stability. Besides, topologically stable configurations are
typically characterized by
 a scale which is  fixed  by the parameters of the Lagrangian, while
configurations discussed in refs.\cite{Anglin,Battye,Ruostekoski}
may posses an  arbitrarily large size. This is another strong hint
of the non-topological nature of configurations considered in
refs.\cite{Anglin,Battye,Ruostekoski}.
\end{thebibliography}
\end{document}